\documentclass[aps,twocolumn,showpacs,showkeys,groupedaddress]{revtex4}

\usepackage{amssymb}
\usepackage{graphicx}
\usepackage{bm}

\makeatletter
\newcommand\erfc{\mathop{\operator@font erfc}\nolimits}
\def\slashchar#1{\setbox0=\hbox{$#1$}
   \dimen0=\wd0 \setbox1=\hbox{/} \dimen1=\wd1
   \ifdim\dimen0>\dimen1 \rlap{\hbox to \dimen0{\hfil/\hfil}} #1
   \else  \rlap{\hbox to \dimen1{\hfil$#1$\hfil}} / \fi}

\makeatother

\begin{document}

\title{Dimension-two gluon condensate from large-$N_c$ Regge models%
\footnote{Research supported by the Polish Ministry of Education and Science,
grants 2P03B~02828 and 2P03B~05925, by the 
Spanish Ministerio de Asuntos Exteriores and the  
Polish Ministry of Education and Science, project 4990/R04/05, 
by the Spanish DGI and FEDER funds with
grant no. FIS2005-00810, Junta de Andaluc\'{\i}a grant No. FQM-225, and
EU RTN Contract CT2002-0311 (EURIDICE)}}

\author{Enrique \surname{Ruiz Arriola}}
\email{earriola@ugr.es}
\affiliation{Departamento de F\'{\i}sica Moderna,
Universidad de Granada, E-18071 Granada, Spain}

\author{Wojciech Broniowski}
\email{Wojciech.Broniowski@ifj.edu.pl}
\affiliation{Institute of Physics, \'Swi\c{e}tokrzyska Academy, PL-25406~Kielce, Poland}
\affiliation{The H. Niewodnicza\'nski Institute of Nuclear Physics, Polish Academy of
Sciences, PL-31342 Krak\'ow, Poland}

\begin{abstract}
It is shown that in the large-$N_c$ limit radial Regge trajectories
give rise in a natural way to the presence of the dimension-2 gluon
condensate, $\langle A^2 \rangle$, in meson correlators. We match
these models to QCD and provide estimates for $\langle A^2 \rangle$ in
terms of other physical quantities. In particular, in the simplest
strictly linear radial Regge model with equal residues $\langle A^2
\rangle$ is proportional to the pion decay constant squared. However,
the linear model fails a consistency condition based on matching the
short- and long-distance string tensions, nor reproduces the
phenomenological values of the gluon condensates.  On the contrary, in
Regge models departing from strict linearity one may reproduce both
the consistency condition and the signs of condensates.
We demonstrate this in a simple explicit model.  
\end{abstract}

\pacs{12.38.Lg, 11.30, 12.38.-t}
\keywords{$\langle A^2 \rangle$ condensate, Regge models, large-$N_c$ limit, 
quark-hadron duality, non-perturbative QCD}
\maketitle

The dimension-two gluon condensate was originally proposed by Celenza
and Shakin~\cite{Celenza:1986th} twenty years ago and appears as rather
elusive gauge-invariant non-perturbative and non-local operator
which generates the lowest $1/Q^2$ power corrections. In fact, in spite
of increasing evidence provided by instanton model
studies~\cite{Hutter:1993sc}, phenomenological QCD sum rules
re-analyses~\cite{Chetyrkin:1998yr}, phenomenological studies of the $\tau$ decay 
data~\cite{Dominguez:1994qt}, theoretical
considerations~\cite{Gubarev:2000eu,Gubarev:2000nz,Kondo:2001nq,Verschelde:2001ia},
quark-model calculations~\cite{Dorokhov:2003kf,Dorokhov:2006ac},
lattice simulations~\cite{Boucaud:2001st,RuizArriola:2004en}, and their
relevance in the confinement-deconfinement phase
transition~\cite{Megias:2005ve}, the dynamical origin of 
$\langle A^2 \rangle$ is still
unclear. For short and recent reviews see,
{\em e.g.}, \cite{Zakharov:2005cg,Narison:2005hb}.

In this paper we stress that within the large-$N_c$ expansion
the $1/Q^2$ corrections appear naturally. The issue was
discussed briefly by Afonin, Andrianov, Andrianov, 
and Espriu~\cite{Afonin:2004yb} in the framework of a Regge model. Here we
reexamine this point and demonstrate that the signs and magnitudes of {\it
both} the dimension-2 and dimension-4 gluon condensates can be
accommodated comfortably with
reasonable values of the parameters of the hadronic spectra. 
The idea of comparing Regge
models to OPE has been explored in a number of
works~\cite{%
Golterman:2001nk,Beane:2001uj,Beane:2001em,%
Simonov:2001di,Golterman:2002mi,Afonin:2003gp}, however
with the exception of Ref.~\cite{Afonin:2003gp} the dimension-2
condensate has been systematically ignored.  We note here a recent
attempt to parameterize such a dimension-2 object within the holographic
approach based on the AdS/CFT correspondence and the resulting Regge 
behavior~\cite{Andreev:2006vy}.

Our method is based on the analysis of the vector and axial vector
meson correlators and has three basic elements: Firstly, we use the
operator product expansion (OPE) of QCD with the non-standard 
$1/Q^2$ power correction in the $V+A$ meson correlator (the
contribution vanishes in the $V-A$ combination). Secondly, confinement
is incorporated in terms of the radial Regge spectra which necessarily
satisfy certain constraints at high energies, in particular, residues
must asymptotically become constant and the string tensions in the
vector and axial vector channels must be the same. Finally, the large
number of colors is assumed, {\em i.e.} the correlators are saturated
with sharp non-interacting meson states. Matching to OPE allows for
enforcement of the two Weinberg sum rules and for the identification
of the QCD condensates in terms of the parameters of the model of the
hadronic spectra. A novel element of our analysis is a consistency 
condition based on matching the values of the short- and long-distance
string tensions.

We begin with the basic formulas in order to fix the notation as well as derive formal
constraints on the Regge models {\em in the presence of the $\langle A^2
\rangle $ condensate}.  
The OPE of the chirally 
even and odd combinations of the transverse parts of the vector and 
axial vector currents, 
$J^{\mu a}_{V,A}=\bar \psi i \gamma^\mu \{ 1,\gamma_5 \} \frac{\tau^a}{2} \psi$, 
gives in the strict chiral limit
\begin{eqnarray}
&&\Pi^T_{V+A} (Q^2) = \frac{1}{4\pi^2} \Big\{-\frac{N_c}{3}
\log \frac{Q^2}{\mu^2} \nonumber \\ &&-
\frac{\alpha_S}{\pi}\frac{\lambda^2}{Q^2} + \frac{\pi}{3}
\frac{\langle \alpha_S G^2 \rangle}{Q^4} + \frac{256 \pi^3}{81}
\frac{\alpha_S \langle \bar q q \rangle^2}{Q^6}\Big\} +\dots \, , \nonumber \\
&& \Pi^T_{V-A} (Q^2) = - \frac{32 \pi }{9}
\frac{\alpha_S \langle \bar q q\rangle^2}{Q^6}+\dots \, , \label{OPE}
\end{eqnarray} 
with the non-standard $1/Q^2$ term proposed in
Ref.~\cite{Narison:2001ix} present. According to
Ref.~\cite{Chetyrkin:1998yr}  the dimension-2 coefficient
$\lambda^2$ is interpreted as the tachyonic gluon mass, $m_g$, providing the
short-distance string tension 
\begin{eqnarray}
\sigma_0 = - 2 \alpha_s \lambda^2/N_c. \label{sigma0}
\end{eqnarray} 
It is also related to the dimension-2 gluon condensate,
$( {3}/{N_c})^2 
\lambda^2 = m_g^2 = - \frac{N_c}{4(N_c^2-1)}g^2 \langle A^2 \rangle$,
where we have inserted appropriate factors of $N_c$, such that 
$\alpha_S \lambda^2 \sim N_c$ and $m_g \sim 1$.

Now comes an important observation: The coefficient of $1/Q^2$ is
proportional to $\sigma_0$, and on the other hand, as we show shortly,
it involves the series over mesonic spectra which depend on the {\em
long-distance} string tension $\sigma$. This allows for building a
sum-rule or a {\em consistency check}, since $\sigma \simeq \sigma_0$.
The near equality is supported by $SU(2)$ lattice
simulations~\cite{Gubarev:2000nz}, where only a $92 \% $ reduction of
$\langle A^2 \rangle $ at the deconfinement transition point is found
suggesting a similar factor between $\sigma$ and $\sigma_0$.

In the large-$N_c$ limit
the vacuum sector of QCD becomes a theory of infinitely many
non-interacting mesons and glueballs, 
hence the correlators 
may be saturated by infinitely
many meson states. Thus one has, 
up to subtractions, 
\begin{eqnarray}
\Pi^T_V(Q^2) &=& \sum_{n=0}^\infty \frac{F_{V,n}^2}{M_{V,n}^2+ Q^2} + c.t., \nonumber \\ 
\Pi^T_A(Q^2) &=& \frac{f^2}{Q^2} + 
\sum_{n=0}^\infty \frac{F_{A,n}^2}{M_{A,n}^2+ Q^2}+c.t., \label{pimes} 
\end{eqnarray} 
where the first term in the axial-vector channel is the massless pion 
contribution with $f=86 {\rm MeV}$ denoting the pion decay constant
in the chiral limit.
Next, we use the radial Regge spectra to saturate 
the vector and axial-vector channels,
\begin{eqnarray}
M^2_{V,n} = M_{V}^2 + a_V n, \; M^2_{A,n} = M_{A}^2 + a_A n, \; n=0,1,\dots \label{regge}   
\end{eqnarray} 
which is well fulfilled~\cite{Anisovich:2000kx} in the experimentally explored
region. 
The vector correlator satisfies the once-subtracted dispersion relation, hence
\begin{eqnarray}
\hspace{-4mm}\Pi^T_V(Q^2) &=& \sum_{n=0}^\infty \left ( \frac{F_{V,n}^2}{M_{V}^2 +
a_V n + Q^2} -  \frac{F_{V,n}^2}{M_{V}^2+a_V n} \right ).
\label{Vsub} 
\end{eqnarray} 
The sum needs to reproduce the $\log Q^2$ term of the OPE expansion (\ref{OPE}), for 
which only the asymptotic part of the spectrum matters. We note that this is the case if
at large $n$ the behavior of the residues is $F_{V,n} \simeq F_V$, {\em i.e.} there
is no $n$-dependence. Similarly, $F_{A,n} \simeq F_A$. If this is satisfied, then 
the sum (\ref{Vsub}) leads to the digamma function, which upon expansion 
produces the $\log Q^2$ term with pure power-law corrections (see the following).
At first sight this seems 
a rather surprising result, since the condition on residues means that all 
highly-excited  radial states are coupled to the current 
with equal strength. However, this is to be expected. The spectral density is    
\mbox{$\rho_V( s ) = F_V^2 \sum_{i=0}^\infty \delta(s-M_{V}^2+a_V n)$},
which from the far Euclidean end looks as a continuum of constant strength in the 
dispersive integral, which
obviously gives rise to the $\log Q^2$. Any dependence of $F_{V,n}$ or 
$F_{A,n}$ on $n$ would spoil
this behavior and damage the twist expansion. Thus the parton-hadron duality
requests, for linear radial Regge spectra, asymptotically constant residues.
  
The chirally-odd combination of currents satisfies the unsubtracted dispersion relation, 
which implies
\begin{eqnarray}
{F_V^2}/{a_V}={F_A^2}/{a_A}. 
\label{equalFoA}
\end{eqnarray}
With this constraint we may now compute the series   
\begin{eqnarray}
&& \hspace{-4mm} \Pi^T_{V-A}(Q^2) 
 \simeq - \frac{F_V^2}{a_V} \psi \left ( \frac{M_{V}^2+Q^2}{a_V} \right )+
\frac{F_A^2}{a_A} \psi \left ( \frac{M_{A}^2+Q^2}{a_A} \right ) \nonumber \\
&&- \frac{f^2}{Q^2} = \frac{F_V^2}{a_V} \log \frac{a_V}{a_A} + {\cal O}(1/Q^2),
\label{Vsub2} 
\end{eqnarray} 
where $\psi(z)=\Gamma'(z)/\Gamma(z)$ is the digamma function, 
and compare to OPE (\ref{OPE}). 
We also assume
$a_V=a_A$, which together with (\ref{equalFoA}) means that 
for asymptotic $n$
\begin{eqnarray}
F_V=F_A=F, \;\;\; a_V=a_A=a.  \label{c1}
\end{eqnarray}
Note that $a_V=a_A$ yields the same density of states in the $V$ and $A$ channels
which complies to the 
``chiral symmetry restoration'' in the spectra \cite{Glozman:2002cp,Glozman:2003bt}.
Moreover, the independent fits of Ref.~\cite{Anisovich:2000kx} 
give very close experimental values $\sqrt{\sigma_A}
= 464 {\rm MeV}$ and $\sqrt{\sigma_V} = 470 {\rm MeV}$, compatible to (\ref{c1}). 

In the first model considered below we thus assume (\ref{regge},\ref{c1}) for all $n$, 
which means strictly linear radial Regge trajectories with constant residues. 
The evaluation of $\Pi^T_{V-A}(Q^2)$ with conditions (\ref{c1}) 
gives
\begin{eqnarray}
&&\hspace{-5mm} \Pi^T_{V-A}(Q^2) = \frac{F^2}{a} 
\left [ - \psi \left ( \frac{M_{V}^2+Q^2}{a}
\right ) +
\psi \left ( \frac{M_{A}^2+Q^2}{a} \right ) \right ] \nonumber \\
&-& \frac{f^2}{Q^2}\,
\simeq \, \Pi^T_{V-A}(Q^2) = \left ( \frac{F^2}{a} (M_{A}^2-M_V^2)-f^2 \right ) 
\frac{1}{Q^2} \nonumber \\ && +
\left ( \frac{F^2}{2a} (M_{A}^2-M_V^2)(a-M_A^2-M_V^2) \right ) 
\frac{1}{Q^4}+ \dots \label{exppl} 
\end{eqnarray}
Matching to (\ref{OPE}) yields the two Weinberg sum rules:
\begin{eqnarray}
 f^2 &=& \frac{F^2}{a} (M_{A}^2-M_V^2), \hspace{27mm} {\rm (WSR~I)} \nonumber \\
 0&=&(M_{A}^2-M_V^2)(a-M_A^2-M_V^2). \hspace{7mm} {\rm (WSR~II)} \nonumber  
\end{eqnarray}

As mentioned, the $V+A$ channel requires regularization. One may use the 
$\zeta$-function, however, a simpler and equivalent procedure is to carry 
the $d/dQ^2$ differentiation, compute the convergent sum, and then integrate
back over $Q^2$. The result is
\begin{eqnarray}
&&\hspace{-5mm} \Pi^T_{V+A}(Q^2) = \frac{F^2}{a} 
\left [ - \psi \left ( \frac{M_{V}^2+Q^2}{a}
\right ) -
\psi \left ( \frac{M_{A}^2+Q^2}{a} \right ) \right ] \nonumber \\
&&+ \frac{f^2}{Q^2} + const \,
\simeq -\frac{2F^2}{a} \log \frac{Q^2}{\mu^2} \nonumber \\ 
&& \hspace{-2mm} + \left ( f^2 +F^2 -\frac{F^2}{a} (M_A^2+M_V^2)   \right ) \frac{1}{Q^2} 
\label{exppl2}\\ && +
\frac{F^2}{6a}\left (a^2 -3 a (M_A^2+M_V^2)+3(M_A^4+M_V^4)  \right ) 
\frac{1}{Q^4}+ \dots \nonumber
\end{eqnarray}
The integration constant has been absorbed in the scale $\mu$.
Matching of the coefficient of the $\log Q^2$ 
to (\ref{OPE}) gives
\begin{eqnarray} 
a= 2\pi \sigma = {24\pi^2} F^2/N_c, \label{c2} 
\end{eqnarray}
where $\sigma$ is the (long-distance) string tension.
If we use $F = 154 {\rm MeV}$ from the
$\rho \to 2 \pi $ decay~\cite{Ecker:1988te} we get $\sqrt{\sigma} = 546 {\rm MeV}$.
The lattice calculation of Ref.~\cite{Kaczmarek:2005ui} gives
$\sqrt{\sigma} = 420 {\rm MeV}$. 
We stress that the conditions (\ref{c1},\ref{c2}) come solely from the asymptotic
spectrum and are insensitive to the low-lying states.
In addition, we may read off from (\ref{exppl2}) the dimension-2 and -4 
condensates. The Regge spectrum assumes confinement, $\sigma \neq 0$, as well 
as $F \neq 0$. The chiral symmetry might a priori be broken or
unbroken. However, since $\sigma \simeq \sigma_0$, also $\sigma_0 \neq 0$, and this 
this via (\ref{sigma0}) implies spontaneous chiral symmetry breaking, $f \neq 0$.
Thus, from WSR~I $M_A \neq M_V$ and 
from WSR~II $a=M_A^2+M_V^2$. In this case (\ref{OPE},\ref{exppl},\ref{exppl2}) give 
\begin{eqnarray}
M_A^2&=&M_V^2+\frac{24\pi^2}{N_c} f^2, \nonumber \\
a&=&M_A^2+M_V^2=2M_V^2+\frac{24\pi^2}{N_c} f^2, \label{cccc2}
\end{eqnarray}
and 
\begin{eqnarray}
-\frac{\alpha_S \lambda^2}{4\pi^3}&=&f^2,\\
\frac{\alpha_S \langle G^2 \rangle}{12\pi}&=&
\frac{M_A^4-4M_V^2 M_A^2 +M_V^4}{48 \pi^2}
\nonumber \\
&=&\frac{288 \pi^4 f^4/N_c^2-24\pi^2 f^2 M_V^2/N_c-M_V^4}{24\pi^2}. \nonumber 
\end{eqnarray}
The numerical value for the dimension-2 condensate is $-\frac{\alpha_S
\lambda^2}{\pi}=0.3 {\rm GeV}^2$ as compared to the value of $0.12
{\rm GeV}^2$ from
Ref.~\cite{Chetyrkin:1998yr,Zakharov:2005cg}. We note that 
Andreev~\cite{Andreev:2006vy} also quotes the estimate $-\frac{\alpha_S
\lambda^2}{\pi}=0.3 {\rm GeV}^2$.
Eq.~(\ref{sigma0}) gives
$\sqrt{\sigma_0} = 782{\rm MeV}$, 
about twice as much as deduced from Eq.~(\ref{c2},\ref{cccc2}), thus the consistency 
check is violated badly. Also, the dimension-4 gluon condensate is
negative for $M_V \ge 0.46$~GeV. Actually it never, not even at
very low values of $M_V$, reaches the QCD sum-rules value of the
condensate.  The dimension-6 condensate 
in the model is zero in the $V+A$ channel, while from OPE it should not
be. All these problems show that the strictly linear radial Regge model
with constant residues is {\em too restrictive}.

\begin{figure}[tb]
\vspace{-13mm}
\begin{center}
\includegraphics[angle=0,width=0.4\textwidth]{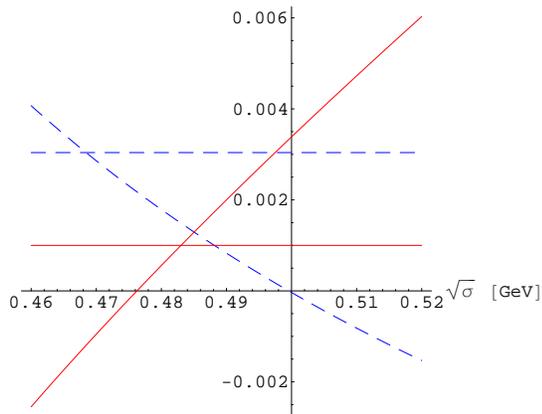}
\end{center}
\vspace{-16mm}
\caption{Dimension-2 (solid line, in GeV$^2$) and dimension-4 (dashed line,
in GeV$^4$) gluon condensates plotted 
as functions of the square root of the string tension. The straight lines 
indicate estimates, see text. The region in $\sqrt{\sigma}$ 
for which both condensates are positive is in the acceptable range compared to 
the estimates of Ref.~\cite{Anisovich:2000kx} and other 
phenomenological and lattice studies.}
\label{fig:gg}
\end{figure}

More parameters can be inserted in models of the spectra by 
treating separately 
the low-lying states, both their residue and position. 
Actually, this is physical. We know that there are departures from the 
linear Regge trajectories at low $n$, also there is no reason why 
at low energies the couplings should be the same.
In principle, these constants are measurable, thus in the large-$N_c$ world 
the values of the OPE condensates can be expressed in terms of the parameters of the 
large-$N_c$ spectra. 
This is very close to the approach of Ref.\cite{Golterman:2001nk,Afonin:2006da}  
with the important extension of admitting the possibility of the dimension-2
condensate. 
For the purpose of illustration we consider the following simple 
modification of the previous model:
\begin{eqnarray}
M_{V,0}&=&m_\rho, \; M_{V,n}^2=M_V^2+a n, \;\; n \ge 1, \nonumber \\
M_{A,n}^2&=&M_A^2+a n, \;\; n \ge 0.
\end{eqnarray}
In words, the lowest $\rho$ mass is shifted, otherwise all is kept ``universal'',
including constant residues for all states. In the present case the 
Weinberg sum rules have the form (we set $N_c=3$ from now on)
\begin{eqnarray}
M_A^2&=&M_V^2+8\pi^2 f^2, \nonumber \\
a&=&8\pi^2 F^2=
\frac{8\pi^2 f^2 \left(4 \pi ^2 f^2+{M_V}^2\right)}{4 \pi ^2 f^2-{m_\rho}^2+{M_V}^2}.
\end{eqnarray}
When $m_\rho=0.77 {\rm GeV}$ is fixed, 
the model has only one free parameter left. We may take it 
to be $M_V$, however, it is more convenient to express it through the string 
tension $\sigma$, 
which is then treated as a free parameter. Thus
\begin{eqnarray}
M_V^2=\frac{-16 \pi ^3 f^4+4 \pi ^2 \sigma f^2-{m_\rho}^2 \sigma}{4 f^2 \pi - \sigma}, 
\end{eqnarray}
and the gluon condensates, obtained by matching to (\ref{OPE}), are
\begin{eqnarray}
-\frac{\alpha_S \lambda^2}{4\pi^3}&=&
\frac{16 \pi ^3 f^4-\pi  \sigma^2+{m_\rho}^2 \sigma}{16 f^2 \pi ^3-4 \pi ^2 \sigma}
\nonumber \\
\frac{\alpha_S \langle G^2 \rangle}{12\pi}&=& 
2 \pi ^2 f^4-\pi \sigma f^2 \nonumber \\ &+&
\frac{3 \sigma \left(\frac{{m_\rho}^2
   \sigma}{\left(\sigma-4 f^2 \pi \right)^2}-2 \pi \right) 
{m_\rho}^2}{8\pi ^2}+\frac{\sigma^2}{12}.
\end{eqnarray}
Figure \ref{fig:gg} shows the two gluon condensates plotted 
as
functions of $\sqrt{\sigma}$.  The constant lines mark the
``physical'' values 
 $-\alpha_s \lambda^2/(4\pi^3)=0.003 {\rm
GeV}^2$ and 
 $\alpha_S \langle G^2 \rangle /(12\pi)=0.001 {\rm
GeV}^4$. Remarkably, the window for which both condensates
are simultaneously positive yields very acceptable values of $\sigma$,
{\em cf.} Ref.~\cite{Anisovich:2000kx}. Moreover, 
the consistency check 
$\sigma =\sigma_0 $ is satisfied for
$\sqrt{\sigma} = 497 {\rm MeV}$. The magnitude of the
condensates is in the ball park of the ``physical'' values.  The value
of $M_V$ in the fiducial range is around $0.82$~GeV -- again a very
reasonable value.  Note the close value of $M_V$ to $m_\rho$,
$0.82$~GeV vs. $0.77$~GeV. This shows sensitivity of the condensates
even to small deviations from the linearity of the Regge
trajectory. 
 The experimental $\rho$ states are at 770, 1450, 1700,
1900, and 2150~MeV, with 
 the last two states not well established,
while the model with $\sigma=(0.47~{\rm GeV})^2$
 gives 770, 1355,
1795, and 2147~MeV. In the $a_1$ channel the experiment 
 shows states
at 1260 and 1640~MeV, while we get somewhat lower 1015 and 1555~MeV.

 Finally, we wish to point out that the $V-A$ channel is actually
very well reproduced 
 with the radial Regge models. As in recent
study of 
 Ref.~\cite{Dorokhov:2003kf}, we apply the Das-Mathur-Okubo
sum rule
to evaluate the low-energy constant $L_{10}$, and the Das-Guralnik-Mathur-Low-Yuong 
sum rule to obtain the electromagnetic pion mass splitting. 
In the strictly linear Regge model with, for instance, $M_A^2 = 2 M_V^2$ and
$M_V=\sqrt{24 \pi^2/N_c} f = 764 {\rm MeV}$, 
we have $a = 3M_V^2$, or $\sqrt{\sigma}= \sqrt{3/2\pi} M_V =
532 {\rm MeV}$, and $F= \sqrt{3} f = 150 {\rm MeV}$,  rather reasonable results. 
Then $ L_{10}=-N_c/( 96 \sqrt{3} \pi)= -5.74 \times 10^{-3} ( -5.5
\pm 0.7 \times 10^{-3})_{\rm exp} $ and $ m^2_{\pi^\pm}-m^2_{\pi_0}
= (31.4 {\rm MeV})^2 \; (35.5 {\rm MeV})^2_{\rm exp}.  $ In our
second model with $\sigma=(0.48 {\rm GeV})^2$ the values are $
L_{10}=-5.2 \times 10^{-3} $ and $ m^2_{\pi^\pm}-m^2_{\pi_0} = (34.4
{\rm MeV})^2,
 $ in quite remarkable agreement with the data.

In conclusion, we note that the scheme of matching OPE to the radial
Regge models 
 produces, in a natural way, the $1/Q^2$ correction to
the vector and axial vector
 correlators, which is attributed to the
dimension-2 gluon condensate.  
 Thus our explicit calculation
illustrates the significance of confinement also for the 
short-distance expansion, as discussed in
Ref.~\cite{Chetyrkin:1998yr,Zakharov:2005cg}. 
 More generally, OPE
with the dimension-2 and all other condensates can be matched 
 by
radial Regge models, provided conditions (\ref{c1},\ref{c2}) are
satisfied by the
 asymptotic spectra, and the parameters of the
low-lying states are adjusted 
 to reproduce the values of the
condensates. In principle, these parameters are measurable, 
 hence
the information encoded in the low-lying states is the same as the
information 
 in the condensates and we could verify consistency. 
Yet the sensitivity of the values of the condensates to the 
parameters of the spectra, as seen by comparing the two
 explicit
models considered in this paper, 
 make such a study difficult at a
more precise level.
 
\begin{acknowledgments}
We thank Oscar Cata, Marteen Golterman, and Santiago Peris for helpful comments.
\end{acknowledgments}


\begin{thebibliography}{30}
\expandafter\ifx\csname natexlab\endcsname\relax\def\natexlab#1{#1}\fi
\expandafter\ifx\csname bibnamefont\endcsname\relax
  \def\bibnamefont#1{#1}\fi
\expandafter\ifx\csname bibfnamefont\endcsname\relax
  \def\bibfnamefont#1{#1}\fi
\expandafter\ifx\csname citenamefont\endcsname\relax
  \def\citenamefont#1{#1}\fi
\expandafter\ifx\csname url\endcsname\relax
  \def\url#1{\texttt{#1}}\fi
\expandafter\ifx\csname urlprefix\endcsname\relax\def\urlprefix{URL }\fi
\providecommand{\bibinfo}[2]{#2}
\providecommand{\eprint}[2][]{\url{#2}}

\bibitem[{\citenamefont{Celenza and Shakin}(1986)}]{Celenza:1986th}
\bibinfo{author}{\bibfnamefont{L.~S.} \bibnamefont{Celenza}} \bibnamefont{and}
  \bibinfo{author}{\bibfnamefont{C.~M.} \bibnamefont{Shakin}},
  \bibinfo{journal}{Phys. Rev.} \textbf{\bibinfo{volume}{D34}},
  \bibinfo{pages}{1591} (\bibinfo{year}{1986}).

\bibitem[{\citenamefont{Hutter}(1993)}]{Hutter:1993sc}
\bibinfo{author}{\bibfnamefont{M.}~\bibnamefont{Hutter}}
  (\bibinfo{year}{1993}), \eprint{hep-ph/9501335}.

\bibitem[{\citenamefont{Chetyrkin et~al.}(1999)\citenamefont{Chetyrkin,
  Narison, and Zakharov}}]{Chetyrkin:1998yr}
\bibinfo{author}{\bibfnamefont{K.~G.} \bibnamefont{Chetyrkin}},
  \bibinfo{author}{\bibfnamefont{S.}~\bibnamefont{Narison}}, \bibnamefont{and}
  \bibinfo{author}{\bibfnamefont{V.~I.} \bibnamefont{Zakharov}},
  \bibinfo{journal}{Nucl. Phys.} \textbf{\bibinfo{volume}{B550}},
  \bibinfo{pages}{353} (\bibinfo{year}{1999}), \eprint{hep-ph/9811275}.

\bibitem[{\citenamefont{Dominguez}(1995)}]{Dominguez:1994qt}
\bibinfo{author}{\bibfnamefont{C.~A.} \bibnamefont{Dominguez}},
  \bibinfo{journal}{Phys. Lett.} \textbf{\bibinfo{volume}{B345}},
  \bibinfo{pages}{291} (\bibinfo{year}{1995}), \eprint{hep-ph/9411331}.

\bibitem[{\citenamefont{Gubarev et~al.}(2001)\citenamefont{Gubarev, Stodolsky,
  and Zakharov}}]{Gubarev:2000eu}
\bibinfo{author}{\bibfnamefont{F.~V.} \bibnamefont{Gubarev}},
  \bibinfo{author}{\bibfnamefont{L.}~\bibnamefont{Stodolsky}},
  \bibnamefont{and} \bibinfo{author}{\bibfnamefont{V.~I.}
  \bibnamefont{Zakharov}}, \bibinfo{journal}{Phys. Rev. Lett.}
  \textbf{\bibinfo{volume}{86}}, \bibinfo{pages}{2220} (\bibinfo{year}{2001}),
  \eprint{hep-ph/0010057}.

\bibitem[{\citenamefont{Gubarev and Zakharov}(2001)}]{Gubarev:2000nz}
\bibinfo{author}{\bibfnamefont{F.~V.} \bibnamefont{Gubarev}} \bibnamefont{and}
  \bibinfo{author}{\bibfnamefont{V.~I.} \bibnamefont{Zakharov}},
  \bibinfo{journal}{Phys. Lett.} \textbf{\bibinfo{volume}{B501}},
  \bibinfo{pages}{28} (\bibinfo{year}{2001}), \eprint{hep-ph/0010096}.

\bibitem[{\citenamefont{Kondo}(2001)}]{Kondo:2001nq}
\bibinfo{author}{\bibfnamefont{K.-I.} \bibnamefont{Kondo}},
  \bibinfo{journal}{Phys. Lett.} \textbf{\bibinfo{volume}{B514}},
  \bibinfo{pages}{335} (\bibinfo{year}{2001}), \eprint{hep-th/0105299}.

\bibitem[{\citenamefont{Verschelde et~al.}(2001)\citenamefont{Verschelde,
  Knecht, Van~Acoleyen, and Vanderkelen}}]{Verschelde:2001ia}
\bibinfo{author}{\bibfnamefont{H.}~\bibnamefont{Verschelde}},
  \bibinfo{author}{\bibfnamefont{K.}~\bibnamefont{Knecht}},
  \bibinfo{author}{\bibfnamefont{K.}~\bibnamefont{Van~Acoleyen}},
  \bibnamefont{and}
  \bibinfo{author}{\bibfnamefont{M.}~\bibnamefont{Vanderkelen}},
  \bibinfo{journal}{Phys. Lett.} \textbf{\bibinfo{volume}{B516}},
  \bibinfo{pages}{307} (\bibinfo{year}{2001}), \eprint{hep-th/0105018}.

\bibitem[{\citenamefont{Dorokhov and Broniowski}(2003)}]{Dorokhov:2003kf}
\bibinfo{author}{\bibfnamefont{A.~E.} \bibnamefont{Dorokhov}} \bibnamefont{and}
  \bibinfo{author}{\bibfnamefont{W.}~\bibnamefont{Broniowski}},
  \bibinfo{journal}{Eur. Phys. J.} \textbf{\bibinfo{volume}{C32}},
  \bibinfo{pages}{79} (\bibinfo{year}{2003}), \eprint{hep-ph/0305037}.

\bibitem[{\citenamefont{Dorokhov}(2006)}]{Dorokhov:2006ac}
\bibinfo{author}{\bibfnamefont{A.~E.} \bibnamefont{Dorokhov}}
  (\bibinfo{year}{2006}), \eprint{hep-ph/0601114}.

\bibitem[{\citenamefont{Boucaud et~al.}(2001)}]{Boucaud:2001st}
\bibinfo{author}{\bibfnamefont{P.}~\bibnamefont{Boucaud}} \bibnamefont{et~al.},
  \bibinfo{journal}{Phys. Rev.} \textbf{\bibinfo{volume}{D63}},
  \bibinfo{pages}{114003} (\bibinfo{year}{2001}), \eprint{hep-ph/0101302}.

\bibitem[{\citenamefont{Ruiz~Arriola et~al.}(2004)\citenamefont{Ruiz~Arriola,
  Bowman, and Broniowski}}]{RuizArriola:2004en}
\bibinfo{author}{\bibfnamefont{E.}~\bibnamefont{Ruiz~Arriola}},
  \bibinfo{author}{\bibfnamefont{P.~O.} \bibnamefont{Bowman}},
  \bibnamefont{and}
  \bibinfo{author}{\bibfnamefont{W.}~\bibnamefont{Broniowski}},
  \bibinfo{journal}{Phys. Rev.} \textbf{\bibinfo{volume}{D70}},
  \bibinfo{pages}{097505} (\bibinfo{year}{2004}), \eprint{hep-ph/0408309}.

\bibitem[{\citenamefont{Megias et~al.}(2006)\citenamefont{Megias, Ruiz~Arriola,
  and Salcedo}}]{Megias:2005ve}
\bibinfo{author}{\bibfnamefont{E.}~\bibnamefont{Megias}},
  \bibinfo{author}{\bibfnamefont{E.}~\bibnamefont{Ruiz~Arriola}},
  \bibnamefont{and} \bibinfo{author}{\bibfnamefont{L.~L.}
  \bibnamefont{Salcedo}}, \bibinfo{journal}{JHEP}
  \textbf{\bibinfo{volume}{01}}, \bibinfo{pages}{073} (\bibinfo{year}{2006}),
  \eprint{hep-ph/0505215}.

\bibitem[{\citenamefont{Zakharov}(2005)}]{Zakharov:2005cg}
\bibinfo{author}{\bibfnamefont{V.~I.} \bibnamefont{Zakharov}}
  (\bibinfo{year}{2005}), \eprint{hep-ph/0509114}.

\bibitem[{\citenamefont{Narison}(2005)}]{Narison:2005hb}
\bibinfo{author}{\bibfnamefont{S.}~\bibnamefont{Narison}}
  (\bibinfo{year}{2005}), \eprint{hep-ph/0508259}.

\bibitem[{\citenamefont{Afonin et~al.}(2004)\citenamefont{Afonin, Andrianov,
  Andrianov, and Espriu}}]{Afonin:2004yb}
\bibinfo{author}{\bibfnamefont{S.~S.} \bibnamefont{Afonin}},
  \bibinfo{author}{\bibfnamefont{A.~A.} \bibnamefont{Andrianov}},
  \bibinfo{author}{\bibfnamefont{V.~A.} \bibnamefont{Andrianov}},
  \bibnamefont{and} \bibinfo{author}{\bibfnamefont{D.}~\bibnamefont{Espriu}},
  \bibinfo{journal}{JHEP} \textbf{\bibinfo{volume}{04}}, \bibinfo{pages}{039}
  (\bibinfo{year}{2004}), \eprint{hep-ph/0403268}.

\bibitem[{\citenamefont{Golterman and Peris}(2001)}]{Golterman:2001nk}
\bibinfo{author}{\bibfnamefont{M.}~\bibnamefont{Golterman}} \bibnamefont{and}
  \bibinfo{author}{\bibfnamefont{S.}~\bibnamefont{Peris}},
  \bibinfo{journal}{JHEP} \textbf{\bibinfo{volume}{01}}, \bibinfo{pages}{028}
  (\bibinfo{year}{2001}), \eprint{hep-ph/0101098}.

\bibitem[{\citenamefont{Beane}(2001{\natexlab{a}})}]{Beane:2001uj}
\bibinfo{author}{\bibfnamefont{S.~R.} \bibnamefont{Beane}},
  \bibinfo{journal}{Phys. Rev.} \textbf{\bibinfo{volume}{D64}},
  \bibinfo{pages}{116010} (\bibinfo{year}{2001}{\natexlab{a}}),
  \eprint{hep-ph/0106022}.

\bibitem[{\citenamefont{Beane}(2001{\natexlab{b}})}]{Beane:2001em}
\bibinfo{author}{\bibfnamefont{S.~R.} \bibnamefont{Beane}},
  \bibinfo{journal}{Phys. Lett.} \textbf{\bibinfo{volume}{B521}},
  \bibinfo{pages}{47} (\bibinfo{year}{2001}{\natexlab{b}}),
  \eprint{hep-ph/0108025}.

\bibitem[{\citenamefont{Simonov}(2002)}]{Simonov:2001di}
\bibinfo{author}{\bibfnamefont{Y.~A.} \bibnamefont{Simonov}},
  \bibinfo{journal}{Phys. Atom. Nucl.} \textbf{\bibinfo{volume}{65}},
  \bibinfo{pages}{135} (\bibinfo{year}{2002}), \eprint{hep-ph/0109081}.

\bibitem[{\citenamefont{Golterman and Peris}(2003)}]{Golterman:2002mi}
\bibinfo{author}{\bibfnamefont{M.}~\bibnamefont{Golterman}} \bibnamefont{and}
  \bibinfo{author}{\bibfnamefont{S.}~\bibnamefont{Peris}},
  \bibinfo{journal}{Phys. Rev.} \textbf{\bibinfo{volume}{D67}},
  \bibinfo{pages}{096001} (\bibinfo{year}{2003}), \eprint{hep-ph/0207060}.

\bibitem[{\citenamefont{Afonin}(2003)}]{Afonin:2003gp}
\bibinfo{author}{\bibfnamefont{S.~S.} \bibnamefont{Afonin}},
  \bibinfo{journal}{Phys. Lett.} \textbf{\bibinfo{volume}{B576}},
  \bibinfo{pages}{122} (\bibinfo{year}{2003}), \eprint{hep-ph/0309337}.

\bibitem[{\citenamefont{Andreev}(2006)}]{Andreev:2006vy}
\bibinfo{author}{\bibfnamefont{O.}~\bibnamefont{Andreev}}
  (\bibinfo{year}{2006}), \eprint{hep-th/0603170}.

\bibitem[{\citenamefont{Narison and Zakharov}(2001)}]{Narison:2001ix}
\bibinfo{author}{\bibfnamefont{S.}~\bibnamefont{Narison}} \bibnamefont{and}
  \bibinfo{author}{\bibfnamefont{V.~I.} \bibnamefont{Zakharov}},
  \bibinfo{journal}{Phys. Lett.} \textbf{\bibinfo{volume}{B522}},
  \bibinfo{pages}{266} (\bibinfo{year}{2001}), \eprint{hep-ph/0110141}.

\bibitem[{\citenamefont{Anisovich et~al.}(2000)\citenamefont{Anisovich,
  Anisovich, and Sarantsev}}]{Anisovich:2000kx}
\bibinfo{author}{\bibfnamefont{A.~V.} \bibnamefont{Anisovich}},
  \bibinfo{author}{\bibfnamefont{V.~V.} \bibnamefont{Anisovich}},
  \bibnamefont{and} \bibinfo{author}{\bibfnamefont{A.~V.}
  \bibnamefont{Sarantsev}}, \bibinfo{journal}{Phys. Rev.}
  \textbf{\bibinfo{volume}{D62}}, \bibinfo{pages}{051502}
  (\bibinfo{year}{2000}), \eprint{hep-ph/0003113}.

\bibitem[{\citenamefont{Glozman}(2002)}]{Glozman:2002cp}
\bibinfo{author}{\bibfnamefont{L.~Y.} \bibnamefont{Glozman}},
  \bibinfo{journal}{Phys. Lett.} \textbf{\bibinfo{volume}{B539}},
  \bibinfo{pages}{257} (\bibinfo{year}{2002}), \eprint{hep-ph/0205072}.

\bibitem[{\citenamefont{Glozman}(2004)}]{Glozman:2003bt}
\bibinfo{author}{\bibfnamefont{L.~Y.} \bibnamefont{Glozman}},
  \bibinfo{journal}{Phys. Lett.} \textbf{\bibinfo{volume}{B587}},
  \bibinfo{pages}{69} (\bibinfo{year}{2004}), \eprint{hep-ph/0312354}.

\bibitem[{\citenamefont{Ecker et~al.}(1989)\citenamefont{Ecker, Gasser, Pich,
  and de~Rafael}}]{Ecker:1988te}
\bibinfo{author}{\bibfnamefont{G.}~\bibnamefont{Ecker}},
  \bibinfo{author}{\bibfnamefont{J.}~\bibnamefont{Gasser}},
  \bibinfo{author}{\bibfnamefont{A.}~\bibnamefont{Pich}}, \bibnamefont{and}
  \bibinfo{author}{\bibfnamefont{E.}~\bibnamefont{de~Rafael}},
  \bibinfo{journal}{Nucl. Phys.} \textbf{\bibinfo{volume}{B321}},
  \bibinfo{pages}{311} (\bibinfo{year}{1989}).

\bibitem[{\citenamefont{Kaczmarek and Zantow}(2005)}]{Kaczmarek:2005ui}
\bibinfo{author}{\bibfnamefont{O.}~\bibnamefont{Kaczmarek}} \bibnamefont{and}
  \bibinfo{author}{\bibfnamefont{F.}~\bibnamefont{Zantow}},
  \bibinfo{journal}{Phys. Rev.} \textbf{\bibinfo{volume}{D71}},
  \bibinfo{pages}{114510} (\bibinfo{year}{2005}), \eprint{hep-lat/0503017}.

\bibitem[{\citenamefont{Afonin and Espriu}(2006)}]{Afonin:2006da}
\bibinfo{author}{\bibfnamefont{S.~S.} \bibnamefont{Afonin}} \bibnamefont{and}
  \bibinfo{author}{\bibfnamefont{D.}~\bibnamefont{Espriu}}
  (\bibinfo{year}{2006}), \eprint{hep-ph/0602219}.

\end{thebibliography}

\end{document}